\documentclass[aps,prl,preprint,nopacs,superscriptaddress]{revtex4}

\usepackage{graphicx}
\usepackage{verbatim}
\usepackage{mathrsfs}
\usepackage{color}
\usepackage{amsmath,amsfonts,amssymb}
\usepackage{graphicx}

\def\3{2.8in}    
\def\2{2.5in}
\def\4{3.0in}

\usepackage{ulem}   
\def \beq {\begin{equation}}
\def \eeq {\end{equation}}
\begin{document}

\title{New fermions on the line in topological symmorphic metals}
\author{Guoqing Chang$^*$}\affiliation{Centre for Advanced 2D Materials and Graphene Research Centre National University of Singapore, 6 Science Drive 2, Singapore 117546}\affiliation{Department of Physics, National University of Singapore, 2 Science Drive 3, Singapore 117542}
\author{Su-Yang Xu\footnote{These authors contributed equally to this work.}$^{\dag}$}\affiliation {Laboratory for Topological Quantum Matter and Spectroscopy (B7), Department of Physics, Princeton University, Princeton, New Jersey 08544, USA}
\author{Shin-Ming Huang}
\affiliation{Department of Physics, National Sun Yat-Sen University, Kaohsiung 804, Taiwan}

\author{Daniel S. Sanchez}\affiliation {Laboratory for Topological Quantum Matter and Spectroscopy (B7), Department of Physics, Princeton University, Princeton, New Jersey 08544, USA}

\author{Chuang-Han Hsu}\affiliation{Centre for Advanced 2D Materials and Graphene Research Centre National University of Singapore, 6 Science Drive 2, Singapore 117546}

\author{Guang Bian}\affiliation {Laboratory for Topological Quantum Matter and Spectroscopy (B7), Department of Physics, Princeton University, Princeton, New Jersey 08544, USA}

\author{Zhi-Ming Yu}\affiliation{School of Physics, Beijing Institute of Technology, Beijing 100081, China} \affiliation {Research Laboratory for Quantum Materials, Singapore University of Technology and Design, Singapore 487372, Singapore}

\author{Ilya Belopolski}\affiliation {Laboratory for Topological Quantum Matter and Spectroscopy (B7), Department of Physics, Princeton University, Princeton, New Jersey 08544, USA}
\author{Nasser Alidoust}\affiliation {Laboratory for Topological Quantum Matter and Spectroscopy (B7), Department of Physics, Princeton University, Princeton, New Jersey 08544, USA}
\author{Hao Zheng}\affiliation {Laboratory for Topological Quantum Matter and Spectroscopy (B7), Department of Physics, Princeton University, Princeton, New Jersey 08544, USA}
\author{Tay-Rong Chang}
\affiliation{Department of Physics, National Tsing Hua University, Hsinchu 30013, Taiwan}
\author{Horng-Tay Jeng}
\affiliation{Department of Physics, National Tsing Hua University, Hsinchu 30013, Taiwan}
\affiliation{Institute of Physics, Academia Sinica, Taipei 11529, Taiwan}
\author{Shengyuan A. Yang}\affiliation{Research Laboratory for Quantum Materials, Singapore University of Technology and Design, Singapore 487372, Singapore}

\author{Titus Neupert}\affiliation {Princeton Center for Theoretical Science, Princeton University, Princeton, New Jersey 08544, USA}

\author{Hsin Lin$^{\dag}$}
\affiliation{Centre for Advanced 2D Materials and Graphene Research Centre National University of Singapore, 6 Science Drive 2, Singapore 117546}
\affiliation{Department of Physics, National University of Singapore, 2 Science Drive 3, Singapore 117542}

\author{M. Zahid Hasan\footnote{Corresponding authors (emails): suyangxu@princeton.edu, nilnish@gmail.com, mzhasan@princeton.edu }}\affiliation {Laboratory for Topological Quantum Matter and Spectroscopy (B7), Department of Physics, Princeton University, Princeton, New Jersey 08544, USA}

\date{\today}

\begin{abstract}

Topological metals and semimetals (TMs) have recently drawn significant interest. These materials give rise to condensed matter realizations of many important concepts in high-energy physics, leading to wide-ranging protected properties in transport and spectroscopic experiments. The most studied TMs, i.e., Weyl and Dirac semimetals, feature quasiparticles that are direct analogues of the textbook elementary particles. Moreover, the TMs known so far can be characterized based on the dimensionality of the band crossing. While Weyl and Dirac semimetals feature zero-dimensional points, the band crossing
of nodal-line semimetals forms a one-dimensional closed loop. In this paper, we identify a TM which breaks the above paradigms. Firstly, the TM features triply-degenerate band crossing in a symmorphic lattice, hence realizing emergent fermionic quasiparticles not present in quantum field theory. Secondly, the band crossing is neither 0D nor 1D. Instead, it consists of two isolated triply-degenerate nodes interconnected by multi-segments of lines with two-fold degeneracy. We present materials candidates. We further show that triply-degenerate band crossings in symmorphic crystals give rise to a Landau level spectrum distinct from the known TMs, suggesting novel magneto-transport responses. Our results open the door for realizing new topological phenomena and fermions including transport anomalies and spectroscopic responses in metallic crystals with nontrivial topology beyond the Weyl/Dirac paradigm.
 \vspace{0.6cm}

\end{abstract}
\maketitle

Understanding nontrivial topology in gapless materials including metals and semimetals has recently emerged as one of the most exciting frontiers in the research of condensed matter physics and materials science \cite{Ashvin_book, TI_book_2015, rev1, Hasan_Na3Bi, Huang2015,Weng2015,Hasan_TaAs, MIT_Weyl,TaAs_Ding,NbAs_Hasan, WT-Weyl,LAG, CA_Ong, CA_Huang, CA_Jia, TaAs_STM_Hasan, TaAs_STM_Yazdani}. Unlike conventional metals, topological metals/semimetals (TMs) are materials whose Fermi surface arises from the degeneracy of conduction and valence bands, which cannot be avoided due to their nontrivial topology. To date, the known TMs include Dirac semimetals, Weyl semimetals, and nodal-line semimetals. Dirac or Weyl semimetals have zero-dimensional (0D) band crossings, i.e., the Dirac or Weyl nodes and a Fermi surface that consists of isolated 0D points in the bulk Brillouin zone (BZ). By contrast, nodal-line semimetals feature one-dimensional (1D) band crossings and a Fermi surface that is made up of 1D closed loops in the BZ. Therefore, the band crossings serve as a key signature of nontrivial topology in metals and can be used to classify TMs. More importantly, these band crossings can give rise to fundamentally new physical phenomena. Since low-energy excitations near the Dirac or Weyl nodes mimic elementary fermions, TMs provide a unique opportunity to study important concepts of high-energy physics such as Dirac fermions, Weyl fermions, and the chiral anomaly in table-top experiments. The correspondence with high-energy physics, in turn, leads to a cornucopia of topologically protected phenomena. The resulting key experimental detectable signatures include the Dirac, Weyl or nodal-line quasiparticles in the bulk, the Fermi arc or drumhead topological surface states on the boundaries, the negative magnetoresistance and nonlocal transport induced by the chiral anomaly \cite{nielsen1983adler,Nonlocal}, the surface-to-surface quantum oscillation due to Fermi arcs \cite{Fermi arc_1,Fermi arc_2}, the Kerr and Faraday rotations in optical experiments \cite{Trivedi}, and topological superconductivity and Majorana fermions when superconductivity is induced via doping or proximity effect \cite{MurakamiNagaosa,Weyl-SC-2, Weyl-SC-7, Weyl-SC-8}. Because all these fascinating properties arise from the band crossings, there has been growing interest in the search for new TMs with new types of band crossings \cite{NF_Kane, NF_Bernevig}, including a classification of 3-, 6-, and 8-fold band degeneracies that appear at high-symmetry points in non-symmorphic crystals~\cite{NF_Bernevig}. Such efforts can lead to new protected phenomena in transport and spectroscopic experiments, which can be potentially utilized in device applications.

In this paper, we identify a class of TMs featuring a type of band crossing beyond the Dirac, Weyl and nodal-line cases. Specifically, we find that the new TM features a pair of triply-degenerate nodes, which are interconnected by multi-segments of lines with two-fold degeneracy. The triply-degenerate node realizes emergent fermionic quasiparticles beyond the Dirac and Weyl fermions in quantum field theory. Moreover, the new band crossing evades the classification of TMs based dimensionality because it is neither 0D nor 1D but rather a hybrid. We show that this band crossing gives rise to a distinct Landau level spectrum, suggesting novel magneto-transport responses. Further, we identify the space groups, in which this new TM state can occur and present material candidates for each space group. Our results highlight the exciting possibilities to realize new particles beyond high-energy textbook examples and to search for new topologically protected low-energy phenomena in transport and spectroscopic experiments beyond the Weyl/Dirac paradigm.

 \vspace{0.4cm}
 
\textbf{Theory of the new band crossing}

\begin{figure}
\includegraphics[width=17cm]{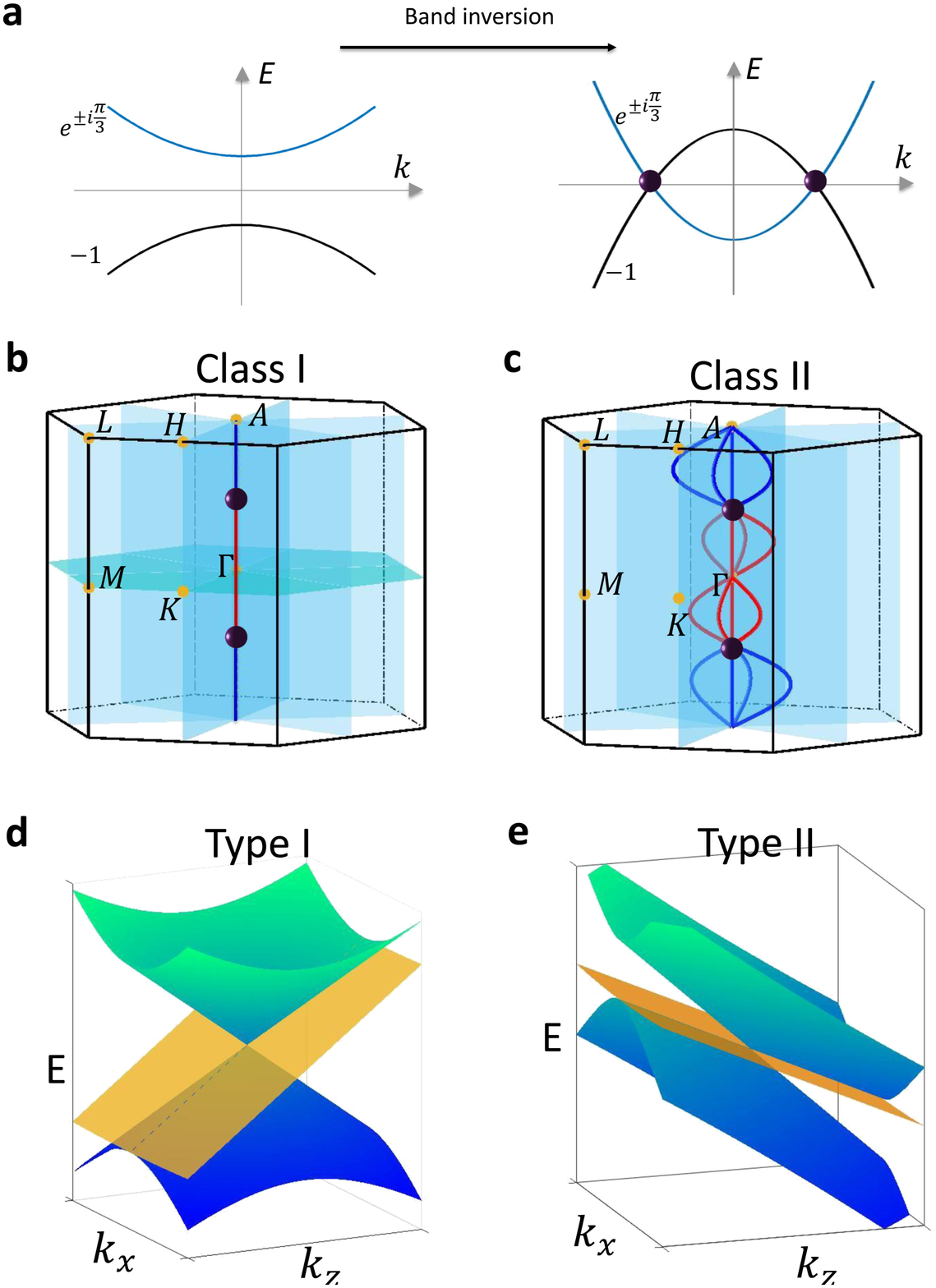}
\caption{\label{Fig1} \textbf{Band crossings in the new topological metal.}  {\bf a,} Cartoon illustration to visualize the band inversion process that drives a three-fold band crossing,}
\end{figure}

We first present a physical picture of the new band crossing without going into mathematical details. We consider an inversion breaking crystal lattice with a three-fold rotational symmetry along the $\hat{z}$ direction ($\tilde{C}_{3z}$) and a mirror symmetry that sends $y\to -y$ ($\tilde{\mathcal{M}}_y$). Note that the $C_{3z}$ rotational symmetry replicates the $\tilde{\mathcal{M}}_y$ twice. In momentum space there are thus in total three mirror planes intersecting along the $k_z$ axis as shown in Figs.~\ref{Fig1}\textbf{a,b}. We first consider the case without spin-orbit coupling (SOC). The $\tilde{C}_{3z}$ operator has three eigenvalues, namely, $e^{-i\frac{2\pi }{3}}$, $e^{i\frac{2\pi }{3}}$, and 1, and we denote the corresponding eigenstates as $\psi_1$, $\psi_2$, and $\psi_3$, respectively. Under the mirror reflection $\tilde{\mathcal{M}}_y$, $\psi_3$ remains unchanged ($\tilde{\mathcal{M}}_y\psi_3=\psi_3$), whereas $\psi_1$ and $\psi_2$ will transform into each other $\tilde{\mathcal{M}}_y\psi_1=\psi_2$; $\tilde{\mathcal{M}}_y\psi_2=\psi_1$. (As an example, consider the $p$-wave basis functions $\psi_1\sim x-iy$, $\psi_2\sim x+iy$.) Thus $\tilde{C}_{3z}$ and $\tilde{\mathcal{M}}_y$ do not commute and cannot be simultaneously diagonalized in the space spanned by  $\psi_1$ and $\psi_2$. Both $\tilde{C}_{3z}$ and $\tilde{\mathcal{M}}_y$ leave every momentum point along the $k_z$ axis invariant. Thus, at each point along the $k_z$ axis, the Bloch states that form a possibly degenerate eigenspace (band) of the Hamiltonian must be invariant under both $C_{3z}$ and $\tilde{\mathcal{M}}_y$. Failure of $\tilde{C}_{3z}$ and $\tilde{\mathcal{M}}_y$ to be simultaneously diagonalizable thus enforces a two-fold band degeneracy of bands with the same eigenvalues as $\psi_1$ and $\psi_2$.
 Therefore, in the absence of SOC, along the $k_z$ axis, the three bands with the three different $\tilde{C}_{3z}$ eigenvalues always appear as a singly-degenerate band ($\psi_3$) and a doubly-degenerate band ($\psi_1$ and $\psi_2$). If the single degenerate and the doubly-degenerate bands cross each other accidentally, a triply-degenerate node will form because their different $\tilde{C}_{3z}$ eigenvalues prohibit hybridization. 
 
When spin is added to the picture, all bands discussed above gain an additional double degeneracy in absence of SOC. However, SOC generically lifts the resulting 6-fold degeneracy into two three-fold degeneracies in absence of inversion symmetry away from the time-reversal symmetric momenta. These three-fold degeneracies are protected for very similar reasons as in the spinless case. The three eigenvalues of the spinful $C_{3z}$ operator are $e^{-i\frac{\pi}{3}}$,  $e^{i\frac{\pi}{3}}$, and $e^{i\pi}$. The same symmetry argument combining $C_{3z}$ and the spinful mirror operator $\mathcal{M}_y$ will show that the two states with $e^{\pm{i}\frac{\pi }{3}}$ eigenvalues must be degenerate. 
Considering all these conditions collectively, the six bands appear as two singly-degenerate bands with the $C_{3z}$ eigenvalue of $e^{i\pi}$ and two doubly-degenerate bands with the $C_{3z}$ eigenvalues of $e^{\pm{i}\frac{\pi}{3}}$. An accidental crossing between a singly-degenerate and a doubly-degenerate band will give rise to a triply-degenerate node along the $k_z$ axis. Away from the $k_z$ axis, all of the three bands can hybridize and the degeneracies will be lifted. Along the $k_z$ axis, a two-fold degenerate nodal line emanates from the three-fold degeneracy, but this degeneracy occurs between the lowest and the middle band on one side of the three-fold degeneracy and between the middle and the highest band on the other side. This structure of degeneracies is reminiscent of but yet distinct from the three-fold degeneracy found for space group 220 in Ref.~\cite{NF_Bernevig}, where pairs of nodal lines emanate from a three-fold degenerate point. The latter is pinned to a high-symmetry point and the symmetries realizing it are quite different from the scenario discussed here.

Now we present the effective Hamiltonian near the triply-degenerate node. In the presence of spin-orbit coupling, we denote three eigenstates of $C_{3z}$ with the eigenvalues of $e^{-i\frac{\pi}{3}}$, $e^{i\frac{\pi}{3}}$, and $e^{i\pi}$ as $\psi_1'$, $\psi_2'$, and $\psi_3'$, respectively. Using the basis $(\psi_1', \psi_2', \psi_3')$, the $C_{3z}$ and $\mathcal{M}_y$ operators have the representations
\begin{equation}
C_{3z}=\left( 
\begin{array}{ccc}
e^{i\frac{\pi }{3}} & 0 & 0 \\ 
0 & e^{-i\frac{\pi }{3}} & 0 \\ 
0 & 0 & -1%
\end{array}%
\right) \text{, }
\qquad
\mathcal{M}_y=\left( 
\begin{array}{ccc}
0 & 1 & 0 \\ 
-1 & 0 & 0 \\ 
0 & 0 & i%
\end{array}%
\right) .
\end{equation}%

It can be seen that $C_{3z}$ and $\mathcal{M}_y$ do not commute with each other, $\psi_1'$ and $\psi_2'$ form a two-dimensional irreducible representation. Therefore, $\psi_1'$ and $\psi_2'$  have to be degenerate at all $k$ points along the $k_z$ axis. We present a $k\cdot{p}$ model for the bands in the vicinity of one triply-degenerate fermion. We denote the momentum relative to the triply-degenerate node as $\boldsymbol{q}=(q_x, q_y, q_z)$. The $k\cdot{p}$ Hamiltonian to linear order in $q_z$  and quadratic order $q_x$ and $q_y$ can be written as
\begin{equation}
H(\boldsymbol{q})=tq_{z}+\left( 
\begin{array}{ccc}
\Delta _{t}q_{z} & -i\lambda q_{+}^{2} & \lambda ^{\prime }q_{+} \\ 
i \lambda q_{-}^{2} & \Delta _{t}q_{z} & i\lambda ^{\prime }q_{-} \\ 
\lambda ^{\prime *}q_{-} & -i\lambda ^{\prime *}q_{+} & -\Delta _{t}q_{z}%
\end{array}%
\right)\text{, }  \label{HTB0}
\end{equation}%
where $q_{\pm}=q_{x}\pm iq_{y}$, the parameters $t$, $\Delta_t$, and $\lambda$ are real, and $\lambda ^{\prime }$ is a complex parameter. The energy eigenvalues are 
\begin{equation}
\varepsilon _{1}=\Delta _{t}q_{z}+\lambda |q_{+ }|^{2}\text{, }\qquad
\varepsilon
_{2,3}=-\frac{1}{2}\lambda |q_{+ }|^{2}\pm \sqrt{2 \lambda ^{\prime 2
}|q_{+ }|^2+\left( \Delta _{t}q_{z}-\frac{1}{2}\lambda |q_{+ }|^{2}\right) ^{2}}\text{,}
\end{equation}%
two of which are degenerate for $q_+=0$.

We now explain how the three-fold band crossing can arise through a band inversion process. Consider a material whose lowest valence and conduction bands are the singly-degenerate band ($\psi_3'$) and the doubly-degenerate band ($\psi_1'$ and $\psi_2'$), respectively. As shown in Fig.~\ref{Fig1}\textbf{a}, if we turn off the hopping of electrons between atomic sites (this can be conceptually done by increasing the lattice constants to infinity), then all bands are flat and the system is an insulator. Now as we gradually increase the magnitude of hopping (this can be conceptually done by decreasing the lattice constant from infinity), bands will gain dispersion. When the band width is large enough relative to the energy offset between the bands, the two bands will be inverted in some interval along the $k_z$ axis (Fig.~\ref{Fig1}\textbf{b}) and cross each other at two points on the opposite sides of the $\Gamma$ point along the $k_z$ axis. These two crossings are the triply-degenerate nodes. This process shows that the triply-degenerate nodes in our new TM always come in pairs and they can move along the $k_z$ axis as the band dispersion is varied.

We find that the new band crossing can be classified into two classes, namely Class I and Class II, depending on whether the mirror symmetry $\mathcal{M}_z$ is present (Class I) or not (Class II). (On the level of the effective Hamiltonian~\eqref{HTB0}, $\mathcal{M}_z$ enforces that $\lambda ^{\prime }$ is real.)
The momentum configurations of band degeneracies in both classes are shown in Figs.~\ref{Fig1}\textbf{c,d}, respectively. They differ in the line degeneracies that connect the triply-degenerate points. In Class I, all band degeneracies are located on the $k_z$ axis. Specifically, two isolated triply-degenerate nodes are located on the opposite sides of the $\Gamma$ point, which arise from the degeneracy between all three ($\psi_1'$, $\psi_2'$ and $\psi_3'$) bands. These two triply-degenerate nodes are linked by non-closed 1D segments with two-fold degeneracy, which arise from the degeneracy between the $\psi_1'$, $\psi_2'$ bands. At any generic $k$ point on the two-fold degenerate segments, the in-plane ($k_x$ or $k_y$) dispersion is a quadratic touching of the $\psi_1'$, $\psi_2'$ bands. The Berry phase along a closed loop around the open segment is $2\pi$, which is trivial. By contrast, in Class II, the two-fold degenerate 1D band crossings form four strands at every cut of constant $k_z$ that join at the triply-degenerate points. At any generic $k$ point on the two-fold degenerate lines, the in-plane ($k_x$ or $k_y$) dispersion is a linear touching of the $\psi_1'$, $\psi_2'$ bands, and the Berry phase around each line is $\pm\pi$, which is nontrivial. One of the four two-fold degenerate segments is pinned to align with the $k_z$ axis. The distinction between Class I and Class II can be understood by an analogy between the Hamiltonian at a generic slice of constant $k_z$ with the Hamiltonian of bilayer graphene: In the latter, the addition of skew interlayer hopping turns a quadratic band touching, corresponding to one degeneracy line segment in Class I, into a quadruplet of Dirac points, corresponding four degeneracy line segments in Class II \cite{BL_Graphene}.

We can further classify the triply-degenerate node by its band dispersion into type-I and type-II, in analogy to a recently introduced notion for Weyl semimetals \cite{WT-Weyl}. In our case, in type-I, the singly-degenerate band and the doubly-degenerate band have Fermi velocities of opposite sign, whereas in type-II all Fermi velocities are of the same sign along the $k_z$ axis. The two situations are separated by a Lifschitz transition. 

We discuss the existence of a topological invariant for the triply-degenerate node. Because a single triply-degenerate node is the end point of a degeneracy line, there is no closed manifold of definite co-dimension, for which a topological invariant can be defined in the usual sense. 
To be specific, taking the topological invariant of a Weyl node (the chiral charge) as an example, it can be calculated by integrating the Berry curvature of the filled bands over a 2D closed manifold in $k$ space that encloses the node. If one tries to calculate a similar integral for the triply-degenerate node, then one would immediately encounter the problem that the line segment degeneracy cuts through the 2D manifold on one side of the triply-degenerate point (e.g. see Fig.~\ref{Fig1}\textbf{e}). In contrast, the topological invariant of a nodal line can be defined as the Berry phase along a closed loop in momentum space. In Class I, any closed loop encircling the line segment that emanates from the triply-degenerate point is contractible (it can be pulled over the degeneracy point) and thus does not yield a topological invariant. In Class II, however, the Berry phase of a loop encircling the line segment that lies on the $k_z$ axis is well defined as this loop is non contractible.
We further note that because the new band crossing discussed here involves a band inversion, one can  calculate the 2D topological invariant on the $k_z=0$ plane, e.g. the $\mathcal{Z}_2$ number or  in Class I the mirror Chern number as done in the topological Dirac semimetal case \cite{Nagaosa}. A 2D topological invariant will guarantee the existence of topological surface states, but such a invariant is not directly related to the triply-degenerate nodes.

 \vspace{0.4cm}
 
\textbf{Band structure and Fermi surface topology of the new topological metal}

In order to understand the band structure of the new topological semimetal, we study its Fermi surface topology. We study the constant energy contours in the ($k_{x}$,$k_{z}$) plane at three different energies $E_{+}$, $E_{0}$, and $E_{-}$, that are above, below, and at the energy of the triply-degenerate node, respectively.  Consider first a pair of type-I triply-degenerate nodes as shown in Figs.~\ref{Fig2}\textbf{a},\textbf{b}. At $E_{+}$, a large electron pocket (T-pocket) enclosing a pair of projected triply-degenerate fermions, and a smaller electron pocket surrounding individual triply-degenerate fermions comprises the Fermi surface. Studying the evolution of both electron pockets as one tunes the binding energy to $E_{-}$ reveals that both pockets shrink, as expected. However, now we observe that the T-pocket (red) is in between the pair of triply-degenerate fermions while a hole pocket (yellow) emerges and surrounds individual triply-degenerate fermions. At the energy of the triply-degenerate fermions, $E_{0}$, only a T-pocket is observed that connects the pair of triply-degenerate fermions. The two observed pockets always have a point of degeneracy, which is due to the two-fold degenerate band along the z-axis. To contrast the Fermi surface behavior of emergent type-I triply-degenerate fermions with those of type-II, we will now study the Fermi surface arising from type-II triply-degenerate fermions, Fig.\ref{Fig2}~\textbf{c}. One clear distinguishing feature in the series of constant energy contours shown in Fig.\ref{Fig2}~\textbf{d} ($E_{+}$ to $E_{-}$) is that there are three Fermi surfaces, which consist of both electron (red) and hole (yellow) pockets at all energies. Furthermore, by scanning through the binding energies, it becomes evident that the electron pocket (T-pocket), composed of two closed contours, encloses the pair projected type-II triply-degenerate fermions at $E_{+}$ and then shrinks to occupy the space in between the pair of triply-degenerate fermions at $E_{-}$. The outer electron pocket is degenerate with the hole pocket at $E_{+}$, which then become disconnected below $E_0$, where now the hole pockets are enclosing the projected type-II triply-degenerate fermions. At energy and momentum space location of the projected type-II triply-degenerate fermions, the three pockets become degenerate, which is consistent with the three-fold and type-II nature of these fermions. 

\vspace{0.4cm}
 
\textbf{Zeeman Coupling} 

In order to understand how the new TM responds to magnetism or magnetic doping in experiments, we study the Zeeman coupling and contrast it with Dirac semimetals. A topological Dirac semimetal system has time-reversal symmetry, space-inversion symmetry, and a uniaxial rotational symmetry along the $k_z$ direction. The presence of time-reversal and space-inversion symmetries requires all bands to be doubly-degenerate because spin up and spin down states have the same energy (Fig.~\ref{Fig3}\textbf{a}). The crossing between two doubly-degenerate bands is realized by a pair of four-fold degenerate points, Dirac nodes, which are protected by the uniaxial rotational symmetry. We consider the effect of a Zeeman field in the $z$ direction, which can be realized by a magnetization or an external magnetic field. Because the Zeeman coupling will lift the spin degeneracy, two doubly-degenerate bands become four singly-degenerate bands. However, since the bands can be distinguished by their rotation eigenvalue, protected two-fold band crossings remain, as shown in Fig.~\ref{Fig3}\textbf{b}. This corresponds to splitting each Dirac node into a pair of Weyl nodes with opposite chiral charge. Each blue shaded area shows the separation between the pair of Weyl nodes that arise from the splitting of a Dirac node in energy and momentum space. These areas also define the regions with non-zero Chern number. Specifically, we consider a 2D $k_x,k_y$ slice of the BZ perpendicular to the $k_z$ axis, and we calculate the Chern number of the band structure on such a slice for all bands below some energy $E$. The Chern number of the slice is only non-zero if the pair $(k_z,E)$ lies within the blue shaded region.

The effect of Zeeman field in $z$ direction, which breaks $\mathcal{M}_y$, is quite different for the new TM. As shown in Fig.~\ref{Fig3}\textbf{c}, the two-fold degeneracy between the $\phi_1'$ and $\phi_2'$ bands is lifted. As a result the doubly-degenerate (blue) band splits into two singly-degenerate bands, each of which crosses with the third band to form a Weyl node. Therefore, each triply-degenerate fermion splits into a pair of Weyl nodes with opposite chiral charge. We point out a number of key distinctions between the Dirac semimetal and the new TM cases. First, in a Dirac semimetal, the immediate pair of Weyl nodes that emerge from the same Dirac node (e.g., $W_1^-$ and $W_1^+$ in Fig.~\ref{Fig1}\textbf{b}) arise from crossings between the same two bands (the yellow and red bands). By contrast, in the new TM, the pair of Weyl nodes that emerge from the same triply-degenerate point (e.g., $W_1^-$ and $W_2^+$ in Fig.~\ref{Fig1}\textbf{d}) arise from the crossings between three different bands. Specifically, $W_1^-$ is due to the crossing between the black and the yellow bands whereas $W_2^+$ is due to the crossing between the yellow and the red bands. As a result, the energy-momentum region with nonzero Chern number (the blue shaded area) in the new TM is drastically different from that of in a Dirac semimetal and spans across all $k_z$. Figures~\ref{Fig3}\textbf{e, f} further show how a triply-degenerate point splits into a pair of Weyl nodes under a Zeeman coupling, for the type-I and -II cases, respectively.

\vspace{0.4cm}
 
\textbf{Landau level spectrum}

In order to understand the magneto transport property of the new TM, we now compare and study the Landau level spectrum arising from triply-degenerate fermions and Weyl fermions. The application of an external magnetic field quantizes the 3D band structure into effective 1D Landau bands that disperse along the $k$-direction that is parallel to the field. In Fig.~\ref{Fig3}\textbf{g} the Landau level spectrum along the $k_{z}$ is shown for a magnetic field applied along the $z$ direction. The Weyl fermion is shown to have a gapless chiral Landau level spectrum. Specifically, besides many parabolic bands away from the Fermi level forming the conduction and valence bands, and we observe a zeroth Landau band (red) extending across the Fermi level. The sign of the velocity of the chiral zeroth Landau level is determined by the chirality of the Weyl fermion. 

This is contrasted with Fig.~\ref{Fig3}\textbf{h}, showing the Landau level spectrum along $k_{z}$ for type-I and type-II triply-degenerate fermions in the left and right panel, respectively. We first point out the similarities between the Weyl fermion and the triply-degenerate fermion cases. We see that the Landau levels found in the Weyl fermion case, i.e., the gapped high Landau levels and the gapless chiral zeroth Landu level, are also observed in the triply-degenerate fermion case. We now emphasize the differences. Firstly, the zeroth Landau level is singly-degenerate in the Weyl fermion case, whereas it is doubly-degenerate in the triply-degenerate fermion case. Secondly, we see a number of additional bands (in blue color) that are roughly parallel to the zeroth chiral Landu level (in red), which do not appear in the Weyl case. We can qualitatively understand these results by visualizing the triply-degenerate band crossing as a Weyl cone plus a third band. This can be clearly seen in the cartoon in Figs.~\ref{Fig1}\textbf{d,e}. For the Landau level structure, the green-blue cone acts like a Weyl cone, while the yellow surface is the third band. They overlap each other on a line that is along the $k_z$ direction. The Landau level spectrum can be explained using this picture. While the Weyl cone will  contribute its characteristic Landau level sturcture, additional Landau levels observed can be explained by the third band. Particularly, if the third band were like a completely flat surface, meaning that it has no dispersion along the in-plane $k_x$ and $k_y$ directions, then all additional bands would be degenerate with the zeroth chiral Landu level, and the zeroth chiral Landau level would have a huge degeneracy. In real materials, the third band will have finite in-plane dispersion. Hence the additional bands become non-degenerate with the zeroth chiral Landau level. This demonstrates that the Landau level spectrum of the triply-degenerate fermion is distinctly different from that of Weyl semimetals. This finding suggests novel magnetotransport responses and further demonstrates the exotic and unique properties of TMs with emergent triply-degenerate fermions.

\vspace{0.4cm}

\textbf{Material realizations}

We have determined the space groups in which the new TM state can occur and identified material candidates for each space group. Importantly, the material candidates that we identified cover both Class I/II and type I/II. The space groups include $\#187$-$\#190$ for Class I and $\#156$-$\#159$ for Class II. A list of the candidate materials is presented in Table I. Here, we take the example of tungsten carbide, WC, as shown in Fig.~\ref{Fig4}. WC crystalizes in a hexagonal Bravais lattice, space group $P$-$6m2$ ($\#187$). The unit cell is shown in Fig.~\ref{Fig4}\textbf{a}), which obviously breaks space-inversion symmetry. The crystal has the $C_{3z}$ rotational symmetry and both horizontal ($\mathcal{M}_z$) and vertical ($\mathcal{M}_y$) mirror planes. Hence we expect the new band crossing to be Class I. Figures~\ref{Fig4}\textbf{c,d} show the first-principles calculated band structures without and with SOC. Triply-degenerate band crossings are seen in both cases. We discuss the band crossing in the presence of SOC in detail. Figure~\ref{Fig4}\textbf{e}, left panel, shows the zoomed-in energy dispersion of the band crossing along $k_z$. It can be seen that the doubly-degenerate band (the blue curve) crosses with two singly-degenerate bands (black curves) forming two triply-degenerate nodes. The right panel shows the in-plane ($k_a$) dispersion that goes through one of the triply-degenerate nodes, where we clearly see that three singly-degenerate bands cross each other at one point. Finally, in Fig.~\ref{Fig4}\textbf{f}, we show that the triply-degenerate nodes in WC indeed split into pairs of Weyl nodes of opposite chirality in the presence of a Zeeman coupling.

\vspace{0.6cm}

In summary, the exploration of TMs has recently experienced a lot of progress and interest. While initially the attraction in TMs was amplified by the realization that the analogues of fermionic particles (e.g. Dirac, Weyl and Majorana fermions) in quantum field theory could be realized in a crystals $k$-space, we are now reaching a point in our understanding that is allowing the study of quasiparticle excitations arising from protected band crossings that do not have a direct analogy in the Standard Model. A crucial insight into the understanding in TMs is the importance of the band crossing dimensionality. While Weyl and Dirac semimetals have zero-dimensional points, the band crossing of nodal-line semimetals forms a one-dimensional closed loop. In this paper, we reported on a new TM that features a triply-degenerate band crossing thereby realizing quasiparticles that have no analog in quantum field theory. Furthermore, the band crossing is neither 0D or 1D, but a combination of both since the two isolated triply-degenerate nodes are interconnected by multiple segments of lines that are doubly-degenerate. We also present a list of crystalline candidate crystals that may realize this new TM. To further elucidate the distinguishing properties of this new three-fold degenerate band degeneracy, we performed detailed calculations on the material candidate WC and studied the Landau level spectrum arising from the node, which is distinct from Dirac and Weyl semimetals. Our results are not only pivotal to the development of our understanding of topological phases of quantum matter, but also provide suitable platforms to experimentally elucidate the transport anomalies and spectroscopic responses in these new TM crystals, which have nontrivial band topology that go beyond the Weyl/Dirac paradigm.

\textit{Note added --- } We remark that the preprints Refs.~\cite{rel-2,rel-1} also study the WC class of materials.

\clearpage
\addtocounter{figure}{-1}
\begin{figure*}[t!]
\caption{which generates a pair of triply-degenerate points (purple spheres). In the weak hopping limit (left panel), electrons can hardly hop from one atomic site to the other. Therefore, the dispersion of bands is very weak. The doubly-degenerate band (blue) and singly-degenerate band (black) are separated by a band gap. As the magnitude of hopping is increased, bands will gain stronger dispersion. When the band width is large enough, the two bands will be inverted in some $k$ region of the BZ and cross each other at two points on the opposite sides of the $\Gamma$ point along the $k_z$ axis. These two crossings are the triply-degenerate nodes. {\bf b, c,} Cartoon showcasing the two distinct classes of the new band crossings: Class I and Class II. The bulk Brillouin zones are represented with the relevant high symmetry points (yellow dots), $k_{z}=0$ mirror plane (turquoise) and three mirror-symmetric planes (blue) along the $C_{3z}$-axis. In Class I, all of the band crossings reside on the $k_{z}$-axis. A pair of triply-degenerate points are connected by non-closed 1D segments with two-fold degeneracy. A trivial $2\pi$ Berry phase was computed along a closed loop around the open segment. In Class II, the two-fold degenerate band crossing form closed contours, which allows for a non-trivial $\pi$ Berry phase to be defined. {\bf d, e} A cartoon illustrating the two types of allowed band dispersions for triply-degenerate nodes: type-I and type-II. Type-I is described by a linear dispersion for both the doubly- and singly-degenerate bands. Type-II is described by band dispersions with the same Fermi velocity direction along $k_{z}$ for both the doubly- and singly-degenerate bands.}
\end{figure*}

\clearpage
\begin{figure}
\includegraphics[width=17cm]{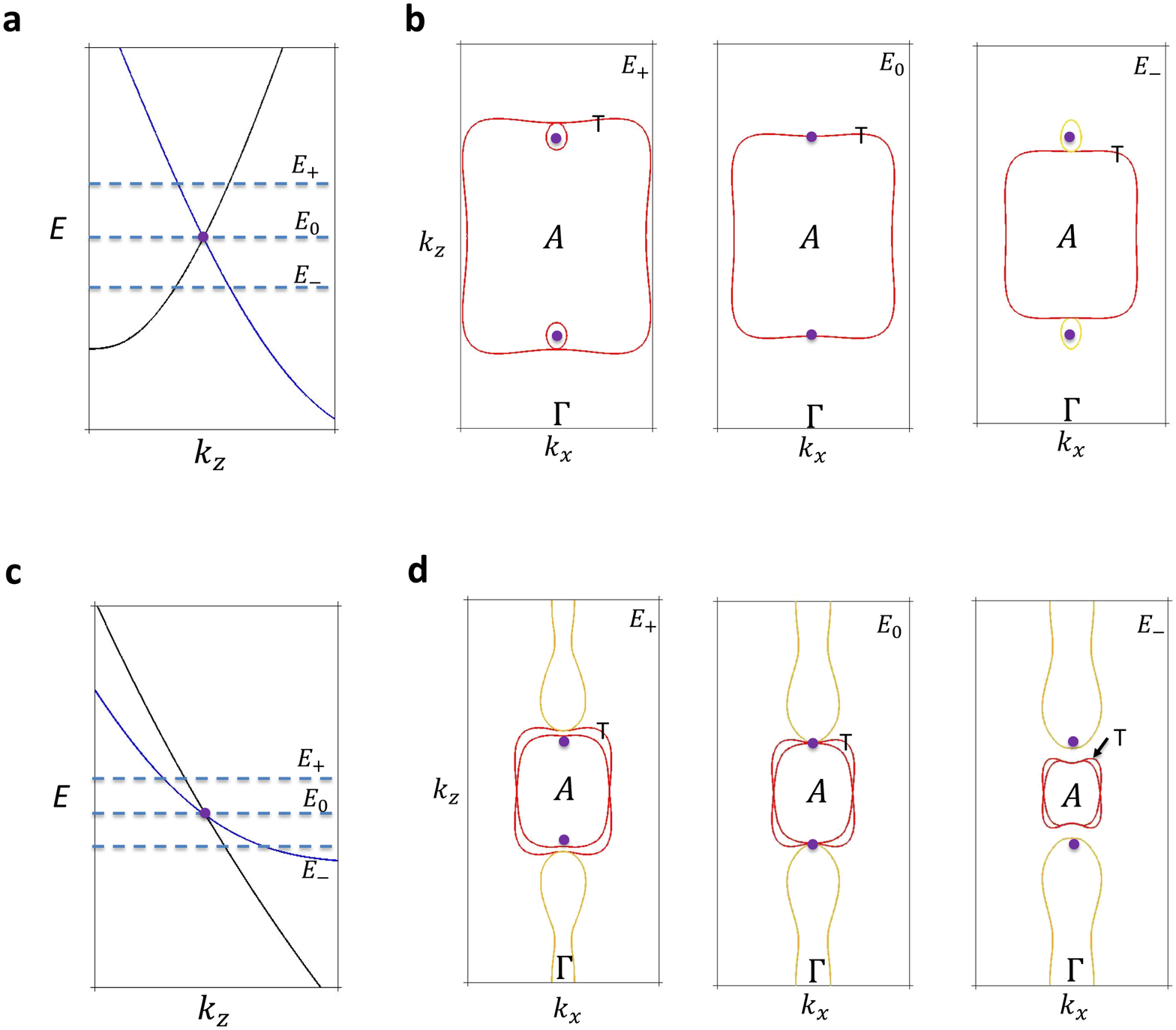}
\caption{\label{Fig2} \textbf{Fermiology of the new topological metal.} {\bf a,} Band dispersion around a type-I triply-degenerate node (purple dot). The binding energies of interest are labeled as $E_{+}$, $E_{0}$ and $E_{-}$ and marked by blue dashed lines. {\bf b,} Constant energy contour calculations for the ($k_{x}$, $k_{z}$) surface. At $E_{0}$ (middle panel), only one Fermi surface (T-pocket) is observed. At $E_{+}$ (left panel) and $E_{-}$ (right panel), two types of contours are observed. Specifically, two electron pockets (red) are observed at E$_{+}$  and one hole pocket (yellow) and one electron pocket is observed at $E_{-}$. Due to the doubly-degenerate band along the $k_{z}$-axis, the two pockets are always degenerate at a point. {\bf c,} Band dispersion around a type-II triply-degenerate node. {\bf d,} Similar to (b) but for (c). In contrast to the two types of constant energy contours observed in (b), (d) clearly shows three types of constant energy contours, which contain both electron-like and hole-like pockets at $E_{+}$, $E_{0}$, and $E_{-}$. At $E_{+}$, two electron pockets (T-pocket) encloses the pair of projected type-II triply-degenerate nodes}
\end{figure}
\addtocounter{figure}{-1}
\begin{figure*}[t!]
\caption{(purple dots) while the hole pocket is degenerate with the outer electron pocket. The calculation at $E_{0}$ (middle panel) reveals that three pockets are degenerate at the location of the projected type-II triply-degenerate fermion. At $E_{-}$ (right panel), the T-pocket is disconnected from the electron pocket and a pair of projected type-II triply-degenerate fermions, while the hole pocket surrounds the projected type-II triply-degenerate fermions.}
\end{figure*}

\clearpage
\begin{figure}
\includegraphics[width=17cm]{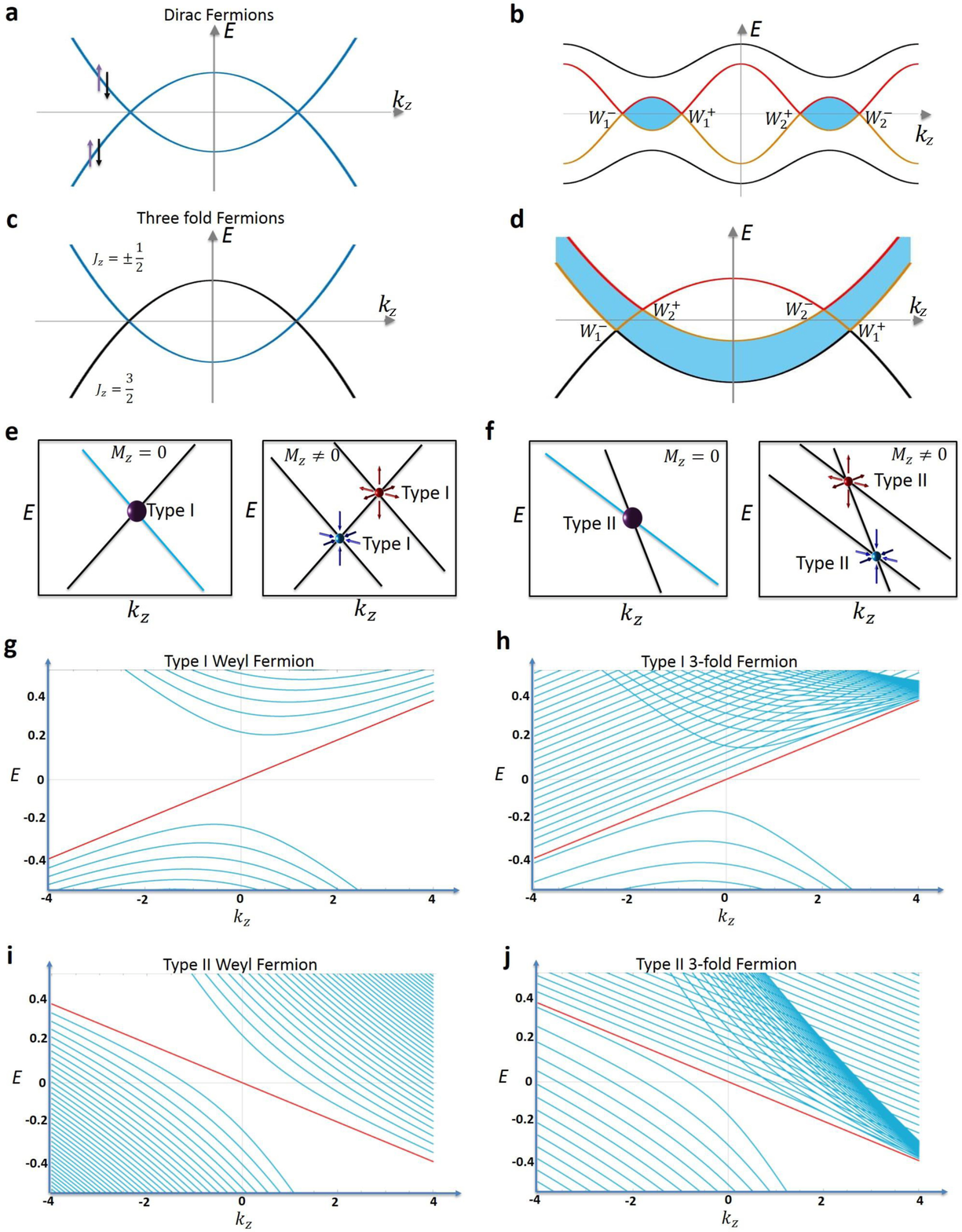}
\caption{\label{Fig3} \textbf{Zeeman coupling and Landau level spectrum.} {\bf a,} Cartoon illustration of two doubly-degenerate bands crossing. The four-fold degenerate crossing point describes Dirac fermion quasiparticles.}
\end{figure}
\addtocounter{figure}{-1}
\begin{figure*}[t!]
\caption{{\bf b,} In the presence of a Zeeman field, the two Dirac fermions described in (a) split into 2 pairs of Weyl fermions (i.e. DP $\rightarrow W^{+}_1 + W^{-}_1$). All pairs of generated Weyl fermions arise from the crossing between two singly-degenerate bands (red and yellow). The blue shaded region corresponds to section in the Brillouin zone with a non-zero Chern number. {\bf c,} Cartoon illustration of a doubly-degenerate band (blue) crossing a singly-degenerate band to create a pair of triply-degenerate fermions at the crossing points. {\bf d,} In the presence of a magnetic field, the doubly-degenerate band splits into two singly-degenerate bands, which then cross the singly-degenerate band at two different locations. Because the pair of generated Weyl fermions are not the crossing points between the same two singly-degenerate bands, the blue shaded area with a non-zero Chern number is distinctly different from the Dirac fermion case presented in (b). {\bf e, f,} A cartoon schematic for the splitting of a triply-degenerate type-I and type-II fermion (purple sphere) in the absence (left panels) and presence (right panels) of a $M_z$ field, respectively. {\bf g, h} Landau level spectrum for type-I (left panel) and type-II (right panel) Weyl fermions, respectively, for a magnetic field along the $z$ direction. Type-I Weyl fermions produce a gapless chiral Landau level spectrum, and realize the chiral anomaly of quantum field theory. Type-II Weyl fermions have a gapless chiral Landau level spectrum only when the magnetic field points along certain directions, and, therefore, realize an anisotropic chiral anomaly. {\bf i, j} Landau level spectrum for Class I triply-degenerate fermions of type-I (left panel) and type-II (right panel), respectively. The zeroth order Landau level band is in red.}
\end{figure*}

\clearpage
\begin{table}[h]
\begin{center}
\begin{tabular}{p{3cm}p{3cm}p{3cm}p{3cm}}
\hline
 Material & Space group  & Type & Class \\
\hline
 WC \cite{WC} & 187 & I and II & I \\
 ZrTe \cite{ZrTe} & 187 & I  & I \\
 $\delta$-TaN \cite{TaN1} & 187 & I and II & I \\
 NbN \cite{NbN} & 187 & I and II & I \\
 VN \cite{VN} & 187 & I and II & I \\
 LiScl$_3$ \cite{LiScI3} & 188 &  II & I \\
 $\epsilon$-TaN \cite{TaN2} & 189 &  II & I \\
 Li$_2$Sb \cite{Li2Sb} & 190 &  II & I \\
 AgAlS$_2$ \cite{AgAlS2} & 156 & I and II & II \\
 AuCd \cite{AuCd} & 157 & I  & II \\
 RuCl$_3$ \cite{RuCl3} & 158 &  II & II \\
 Ge$_3$N$_4$ \cite{Ge3N4}& 159 & I and II & II \\
\hline
\end{tabular}
\end{center}
\caption{\textbf{A list of candidates for the new topological metal.}.}
\end{table}

\clearpage
\begin{figure}
\includegraphics[width=17cm]{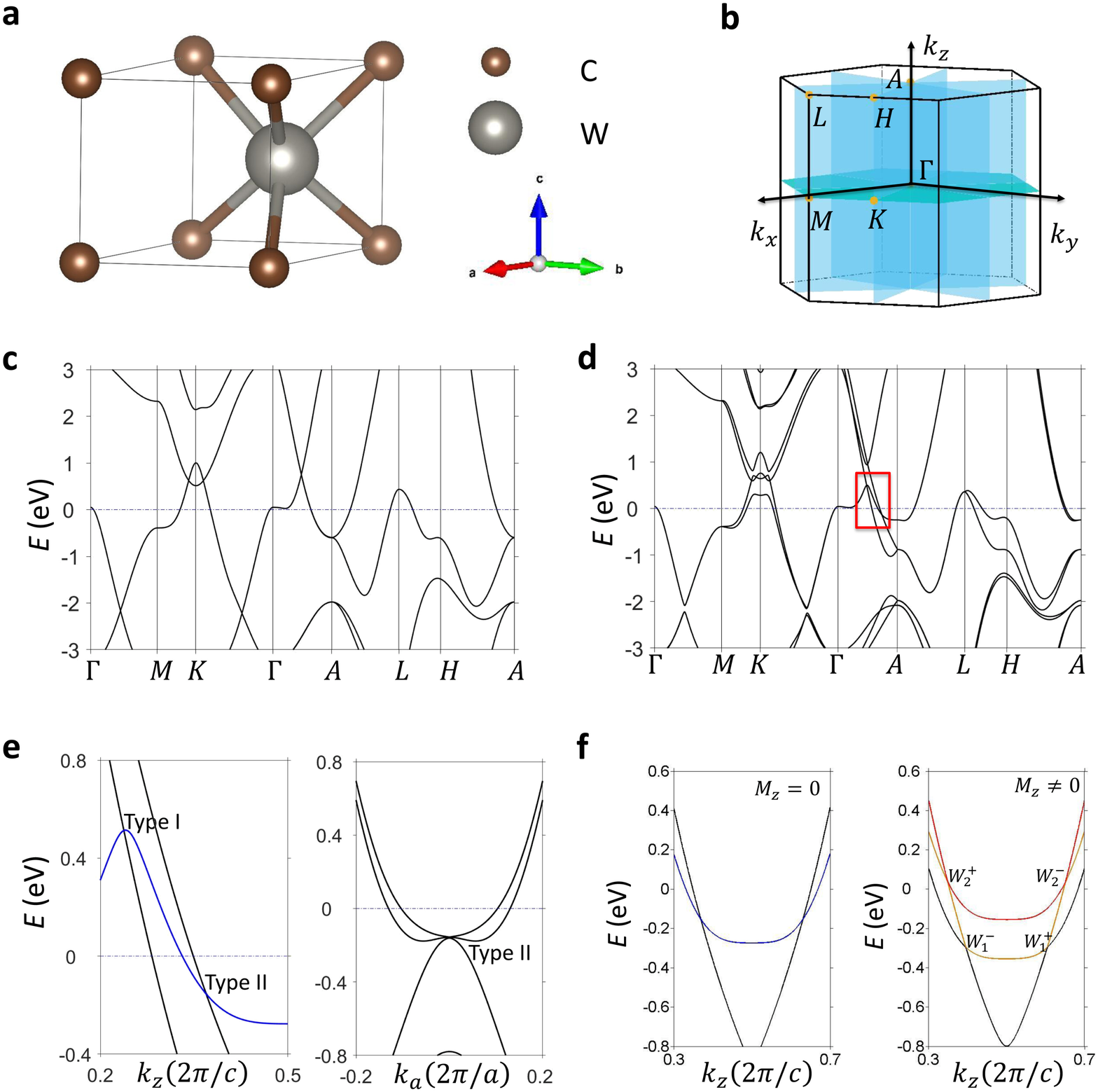}
\caption{\label{Fig4} \textbf{Material realizations of the new topological metal in WC class of materials} {\bf a,} Crystal structure of WC with space group $P$-$6m2$ ($\#187$), showing the W and C atoms as silver and bronze spheres. {\bf b,} The corresponding bulk Brillouin zone with the relevant high symmetry points (yellow dots), $k_{z}=0$ mirror plane (turquoise), and three mirror planes (blue) that intersect along the $C_3$-axis. {\bf c,} Band structure calculation of WC without SOC. In the absence of SOC, the crossing along $M-K-\Gamma$ results in a nodal ring around the $K$ point. {\bf d,} Same calculation as in (c) but with the inclusion of SOC. Enclosed in the red rectangular box are two observed crossings points along the $\Gamma-A$ line. Furthermore, inclusion of SOC allows for the touching points along $M-K-\Gamma$ that are protected by the $k_{z}=0$ mirror plane to remain and form two nodal rings around the $K$ point.}
\end{figure}
\addtocounter{figure}{-1}
\begin{figure*}[t!]
\caption{{\bf e,} In the left panel, a zoomed-in calculation of the region within the red rectangular box in (d) reveals that the doubly (blue) and singly (black) degenerate bands cross at two different energies. The triply-degenerate node above the Fermi level is type-I, and the one below the Fermi level is type-II. In the right panel, the type-II character of the triply-degenerate fermion is shown by cutting through the degeneracy point along the $k_{a}$-direction. {\bf f,} Zoomed-in calculation of the observed type-II triply-degenerate crossing in (e) in the absence (left panel) and presence of a magnetic field along the $k_{z}$-direction. The application of the field along this direction preserved $C_{3}$ symmetry, and results in the triply-degenerate fermion to split into a pair of Weyl fermions by splitting the doubly-degenerate band into two singly-degenerate bands. The resulting two Weyl fermions are labeled as $W_1$ and $W_2$, marking the crossing points between the black/yellow and red/yellow bands, respectively.}
\end{figure*}

\end{document}